\documentstyle[twocolumn,aps,psfig]{revtex}
\begin{document}
\draft

\twocolumn[\hsize\textwidth\columnwidth\hsize\csname @twocolumnfalse\endcsname

\title{ Local Enhancement of Antiferromagnetic Correlations by
Nonmagnetic Impurities}

\author{George Balster Martins$^1$, Markus Laukamp$^1$, Jos\'e Riera$^2$, and Elbio Dagotto$^1$}
\address{$^1$ National High Magnetic Field Lab and Department of Physics, Florida State University, Tallahassee,\\
Florida 32306,USA}
\address{$^2$ Instituto de F\'{\i}sica Rosario, Avenida 27 de Febrero 210 bis, 2000 Rosario,\\
Argentina}
\maketitle

\begin{abstract}

The local enhancement of antiferromagnetic correlations near vacancies
observed  in a variety of spin systems
is analyzed in a single framework. Variational calculations suggest
 that the resonating-valence-bond 
character of the spin correlations at short distances        
is responsible for the
enhancement. Numerical results for uniform
spin chains, with and without frustration, dimerized chains, 
ladders, and two dimensional
clusters are in agreement with our conjecture. This
short distance phenomenon occurs independently of the long distance behavior of
the spin correlations in the undoped system. Experimental predictions
for a variety of compounds are briefly discussed.

\end{abstract}


\vskip2pc]
\narrowtext

Studies of ladder compounds continue producing
fascinating results. In addition to the discovery of
a spin gap in undoped even-leg ladders\cite{takano}, 
superconductivity at high pressure  
in ${\rm Sr_{0.4} Ca_{13.6} Cu_{24} O_{41.84} }$,
with 2-leg ladders and chains in its structure,
has been recently reported\cite{akimitsu}. 
Both properties, 
predicted by theoretical arguments,\cite{ladder} indicate a close 
interplay between the spin and charge degrees of freedom 
leading to a rich phase diagram. More recently, the doping
of ladders with nonmagnetic impurities (replacing 
spin 1/2 ${\rm Cu^{2+}}$ by spin 0 ${\rm Zn^{2+}}$)
has revealed another surprising property:
the spin gap is rapidly suppressed as the Zn concentration 
increases, and an antiferromagnetic (AF) phase is
stabilized\cite{azuma}. A
 similar behavior has also been  observed in 
spin-Peierls chains\cite{hase1}, which have a spin gap produced by
dimerization. 
The phenomenon  is interesting since a spin ordered state
is generated by the random replacement of spins by vacancies, 
an apparently disordering procedure. 
These results have been recently 
addressed with one dimensional (1D) spin models using field
theory\cite{fukuyama} and numerical techniques.
Computational studies
found that the AF correlations near a vacancy in dimerized
chains\cite{recent1d} and 2-leg ladders\cite{recent1d,recentladder} 
are  enhanced with respect to the undoped case.
It was conjectured that this local enhancement may trigger the 3D AF
order in Zn-doped dimerized chains and ladders.
In-gap weakly interacting
$S=1/2$ localized states were found near Zn\cite{sandvik}.
However, the microscopic origin of the local AF enhancement near a vacancy
is still not intuitively understood.

Independently of these recent developments, related phenomena have
been discussed in a variety of contexts: 1.
A staggered moment appears near a vacancy for 1D
$S=1$ Heisenberg systems\cite{sorensen}; 
2. The undimerized 1D
$S=1/2$ Heisenberg model has 
an enhanced spin structure factor $S(\pi)$ near
vacancies according to boundary conformal field theory and
Monte Carlo (MC) simulations\cite{eggert}; 3. 
Near a vacancy injected into a 2D N\'eel ordered state, the staggered
moment increases with respect to the undoped system\cite{bulut}.

In this paper it is proposed
that all these examples of locally 
enhanced antiferromagnetism near a vacancy, which have been studied 
independently in the literature, may have a 
simple common explanation.
The unifying picture relies on the
resonating-valence-bond (RVB) 
character of the spin correlations at short distances
for a variety of Heisenberg
 spin systems where the  nearest-neighbors (NN) 
interaction, regulated by a
coupling $J_1$, is the largest. 
Independently of the long distance properties of the 
model, the small distance behavior, at 
least for small spin $S$,  is dominated by the formation
of short spin singlets resonating in all their possible
arrangements\cite{anderson}.
Actually, RVB variational states have
large overlaps with exact ground states of finite clusters, even for
2D N\'eel ordered systems\cite{overlap}. 
Corrections involve long singlets and triplets
with a weight decaying
as the spin-spin distance grows.

In the case of a 1D 
$S=1/2$ chain, Fig.1a-b illustrates two  arrangements of spin singlets
that carry important weight in the undoped ground state. 
Consider  now a vacancy introduced at
site $0$, effectively cutting  the chain if a NN spin
Hamiltonian is used.
In this situation configuration Fig.1b is replaced  by Fig.1c that
contains a ``free'' spin. The NN exchange 
strongly favors the coupling of 
this free spin with the spin at site 2. Then,
configuration Fig.1d is equally probable, and the argument can be 
repeated moving the free spin over large distances 
(spin-charge separation). The final dominant
configuration near a vacancy has the form shown 
in Fig.1e. The key detail to understand the AF enhancement 
is that in the absence of the vacancy the
spin at 1 spends roughly half the time coupled into a singlet with the
spin at 2 and the other half with the spin at 0, while with a
vacancy the spin at 1 forms a singlet most of the time with 
the spin at 2. Then, the 1-2 spin correlation is now 
enhanced. The vacancy 
has $pruned$ the possible singlet configurations and the spins next to it
no longer ``resonate'', but their singlet partners are fixed by
geometry.

\begin{figure}[htbp]
\centerline{\psfig{figure=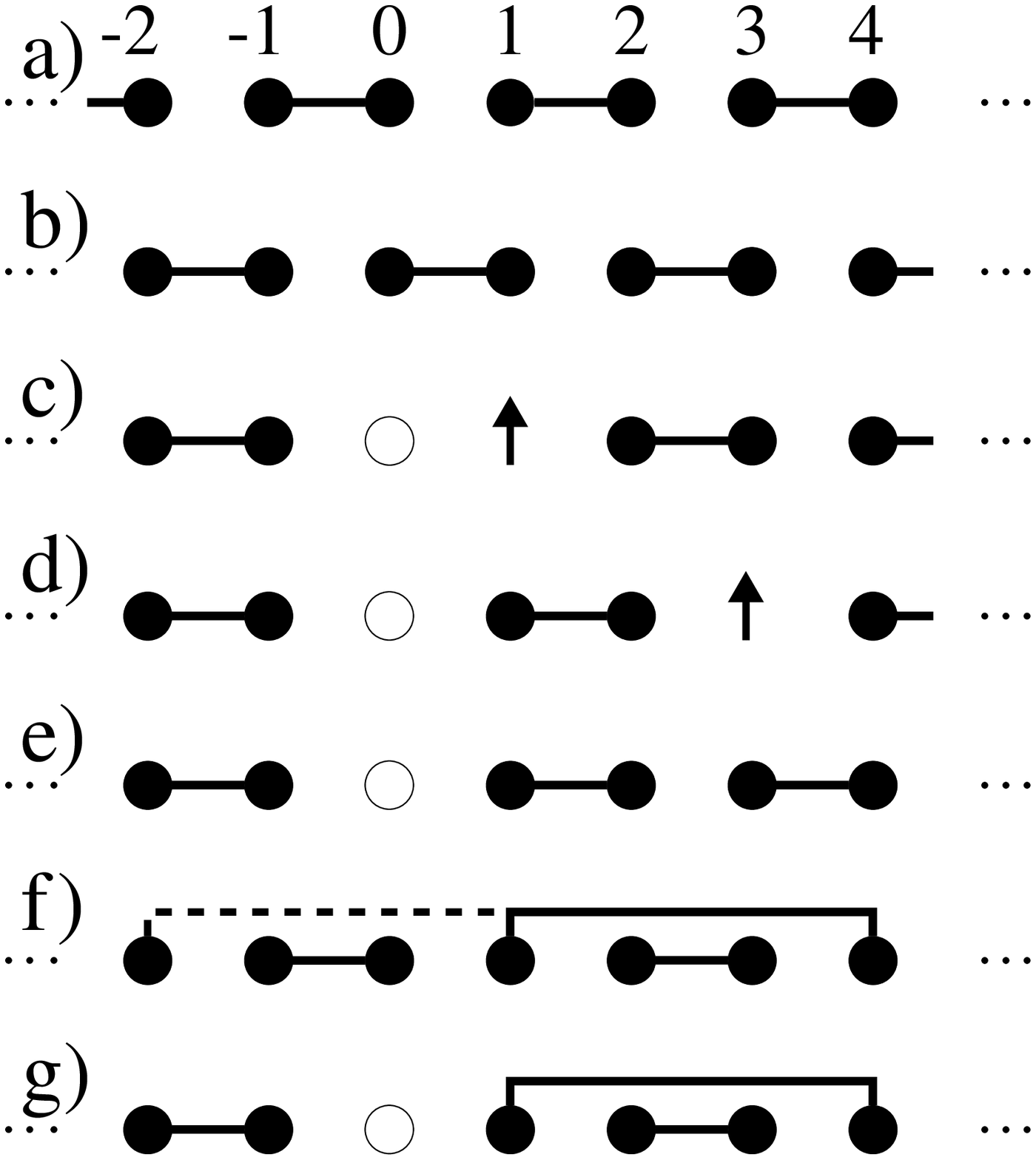,height=5cm}
\psfig{figure=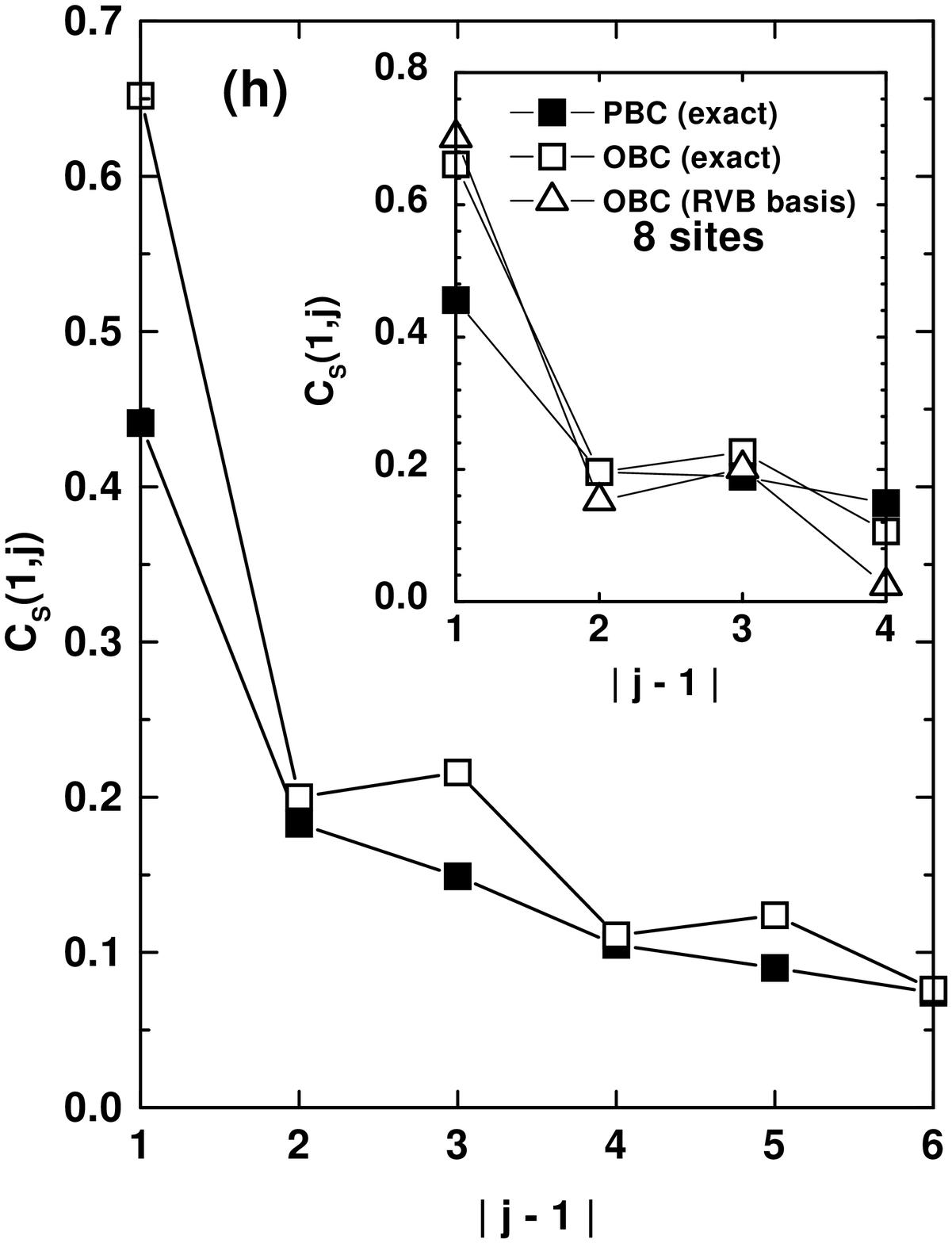,height=6cm}}
\vspace{0.5cm}
\caption{(a)-(g) Examples of spin singlets relevant for our
discussion (see text); (h) $C_S(1,j)$ for a $S=1/2$
Heisenberg model on a 128 sites chain calculated with DMRG. 
The open (full) squares are results with (without) a
vacancy at site 0. The inset contains the RVB variational results (see
text) for 8-sites with OBC compared against exact results for
the same chain with OBC and PBC. 1 is the first site of the chain.}
\end{figure}

The same pruning of RVB states occurs when 
corrections to Fig.1a-b are included.
Consider a configuration with a singlet linking sites 1 and 4 (Fig.1f).
Such configuration produces the largest contribution to the spin
correlation at distance 3.
As in the previous case, the spin at 1 divides its 
tendency to form singlets equally
between sites 4 and -2. But when a vacancy
is introduced, only one possibility
for the coupling of the spin  at 1 with partners at distance 3
remains (Fig.1g), enhancing the correlation at that  distance.
Again the nonmagnetic impurity has pruned the singlet configurations 
increasing the AF correlations in its vicinity. The same
argument applies for longer singlets, 
although moving away from the vacancy  the AF enhancement is 
reduced since eventually both
Fig.1a-b must dominate in the bulk. Results 
for the ground state staggered spin correlation
$C_S(i,j) = \langle { {{\bf S}_i}\cdot{{\bf S}_j}} \rangle
(-1)^{i+j}$, in the
standard notation, calculated with 
density matrix renormalization group (DMRG) techniques\cite{white}
are shown in Fig.1h for the spin 1/2 Heisenberg
chain. The AF enhancement is large 
($48\%$ for the bond next to the vacancy).
The effect can be rephrased as an increase in $S(\pi)$ as
previously observed by Eggert and Affleck\cite{eggert}.
We here confirm their results and provide
a simple explanation for the origin of the enhancement.

To provide further support to these ideas, a RVB variational
calculation on a chain of 8 sites with open boundary conditions (OBC) was
performed. The state used is a linear
combination of (a) the state with four NN singlets, (b) all states with
three NN singlets and one singlet of length 3,
and (c) all states with two triplets of length 2 and two NN singlets.
The relative weights are fixed minimizing the energy.
The spin correlations in this variational state are also
shown in Fig.1h (inset).  Contrasting the results against exact
calculations with OBC, it is clear that
such a simple state contains the main 
features observed in the actual ground state of longer chains. 

The discussion can be easily extended to NN $S=1/2$ Heisenberg
models with a larger coordination number.
Consider first a 2-leg ladder.
Snapshots of the spin arrangement in the vicinity of an arbitrary
site, e.g. 1,
would show that this spin spends most of its time singlet coupled 
with spins 2,3 and 4 (Fig.2a)\cite{comment2}.
When a vacancy is introduced at site 2, now 
the spin at 1
forms singlets only with two partners rather than three (Fig.2b).
This pruning of the RVB basis 
enhances the spin correlation at distance 1 along the chain, as in 1D.
Fig.2d contains $C_S(i,j)$ for a 2-leg ladder. Shown are
the DMRG spin correlations  both along the leg where the vacancy is,
and in the opposite leg. The results are in good agreement with
previous calculations\cite{recentladder}.
The enhancement is substantial only  for same-leg correlations, 
and even in this case it is smaller than for the $S=1/2$
chain, in agreement with the picture discussed before 
(i.e. counting links the naive
enhancement ratio would be 3/2 for ladders vs
2/1 for chains). At distance
1 and for same-leg correlations, the
actual enhancement is $\sim 22\%$.

\begin{figure}[htbp]
\centerline{\psfig{figure=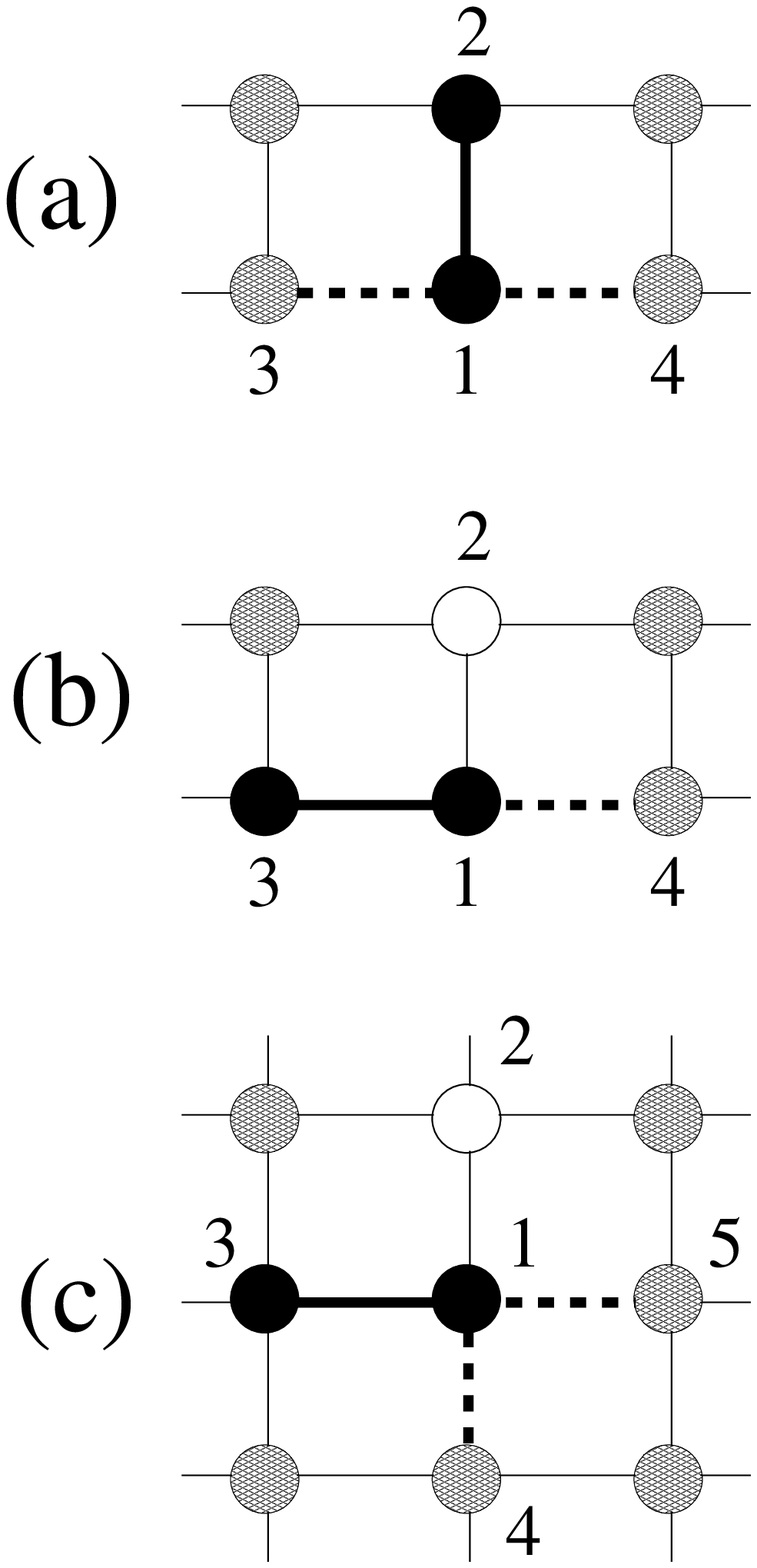,height=7cm}\psfig{figure=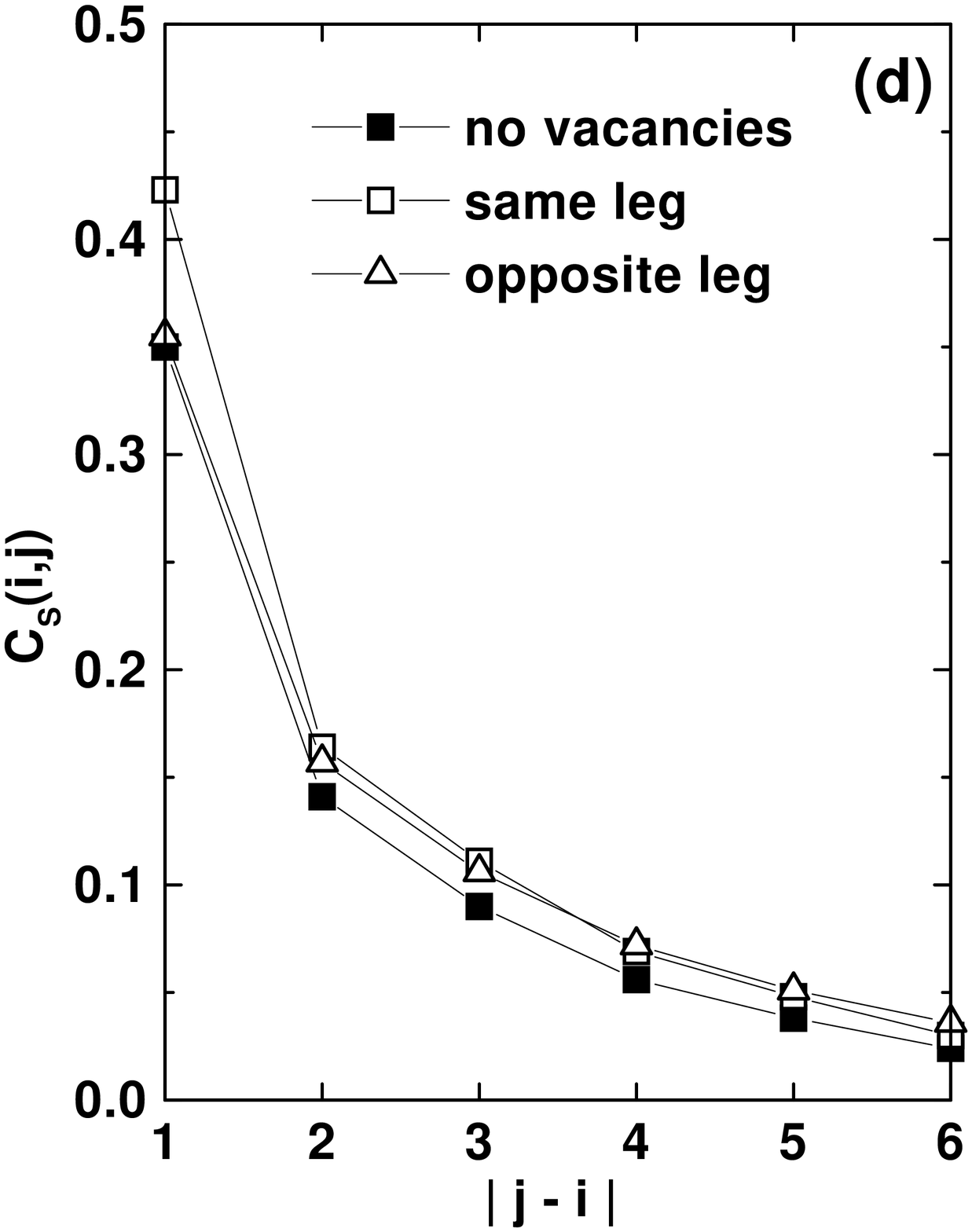,height=7cm}}
\vspace{0.5cm}
\caption{(a)-(c) Spin singlets relevant for ladders and hypercubic
systems (see  text);
(d) $C_S(i,j)$ for a 2-leg ladder calculated with DMRG
on a  $2 \times 32$ cluster.
The open squares (triangles) are spin correlations
along the same (opposite) leg where the vacancy is located, with
a starting site $i$ next to the vacancy. The
full squares are results without
vacancies. The vacancy is located at the center of the cluster.}
\end{figure}

Numerically the short distance spin correlations
smoothly interpolate between the results of
chains and 2-leg ladders 
as the  rung coupling $J_{\perp}$ is reduced with respect
to the chain coupling  $J$. 
As example, in Fig.3a exact diagonalization (ED)\cite{review}
results are shown for $J_{\perp}/J=0.5$ using periodic boundary
conditions
(PBC) and OBC.
They suggest that the short distance
AF enhancement of chains and ladders have a common
origin independent of the 
long distance behavior  of their spin correlations (power-law vs
exponential decay). Only the range of the Zn-induced disturbance
is affected by the presence of a spin gap.
For example,
due to spin-charge separation in 1D, the $S=1/2$ state
obtained by removing one spin from an even site chain is
uniformly distributed over the chain\cite{sandvik}. However, near the
vacancy the AF correlations are nevertheless enhanced (Fig.1h).
Thus, the short distance effect discussed here does 
not seem correlated with the presence of a $S=1/2$ 
localized state near the vacancy. 
To further confirm these ideas, correlations
for a $J_1-J_2$ chain  
($J_2$ being a next to NN coupling) were calculated 
varying $J_2/J_1$ from 0 (no gap in the ground state)
to the exactly solvable point
0.5 (dimerized gapped state).
A smooth interpolation between the two regimes was observed at short
distances (see e.g. Fig.3b that includes a RVB variational 
calculation). 
As the Zn impurity density
grows, the short distance effects  will
eventually  govern the behavior of 
the system, and the tail in the spin disturbance
becomes irrelevant.

The analysis of models for CuGeO$_3$ require a special discussion.
Here a dimerization occurs at low temperatures\cite{hase1}
which  will be represented by a modulation of the original NN
exchange using $J(1 \pm \delta) {{{\bf S}_{ i}}\cdot{ {\bf S}_{
j}}}$, with $\delta \sim 0.03$\cite{riera1}. Note that a
 more proper description
of CuGeO$_3$ requires the inclusion of phonons to generate the
dimerization dynamically. However, the
model used here is sufficient for our qualitative studies.
The $+$($-$) sign corresponds to ``strong'' (``weak'') links.
The addition of a next NN interaction $J_2/J_1 \sim 0.2-0.3$ is also 
needed in CuGeO$_3$\cite{riera1}.
Based on the RVB picture discussed before it would not be surprising
 that AF
enhancements of short distance spin correlations will also appear 
in Zn-doped spin-Peierls chains. In fact we found the surprising
effect that the enhancement in this case
is the largest among the family of
models studied here. As example, consider the results of Fig.3c,
obtained exactly on a finite chain 
with OBC. The pattern of strong vs weak
links considered here 
is such that sites 1 and 2 next to the end are connected by a weak link. 
Fig.3c shows a large $\sim 84 \%$
enhancement of $C_S(1,2)$
compared with results for a ring. The reason is that for a ring
spins 1 and 2 are not much
correlated since each can form a strong bond with a neighboring spin, 
while for an open chain  the spin at 1 is
``free'' and strongly tries to form a singlet with its neighbor 
regardless of the smaller value of the coupling $J(1-\delta)$ in this
link. Thus, the relative enhancement is large. If the pattern of weak
vs strong bonds 
is shifted in one lattice spacing, then the AF enhancement 
is smaller since the spins of the now 
strong bond 1-2 are strongly
correlated for both a ring and a chain with OBC.

\begin{figure}[htbp]
\centerline{\psfig{figure=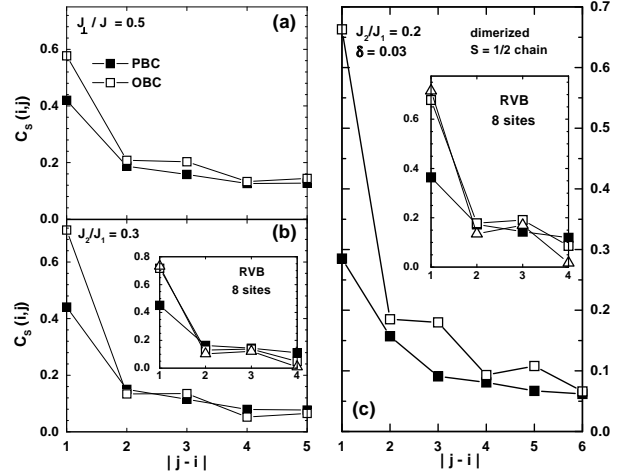,height=7cm}}
\vspace{0.5cm}
\caption{ $C_S(i,j)$ obtained with ED for (a) a 2-leg ladder with $J_{\perp}/J=0.5$
using PBC (full squares) and OBC (open squares). The correlations
are measured from the edge of  
a $2 \times 10$ cluster; (b) a
$S=1/2$ chain with $J_2/J_1=0.3$ and 14 sites. Open squares
denote results from the end of an open chain, while full squares are
results for a ring. The inset shows RVB
variational results with the same state used in  Fig.1h. Open (full)
squares are exact results with open (periodic) boundary conditions.
The triangles denote the variational predictions;
(c) a dimerized $S=1/2$ chain with OBC, 14 sites, and
$J_2/J_1 = 0.2$, $\delta = 0.03$. {\it i} is the first spin on the chain.
The inset contains RVB
results on an 8 site open chain. The symbols are as in Fig.3b.
The first link in the chain near the edge is ``weak''.}
\end{figure}

The local RVB-picture presented here predicts that AF enhancement
should also occur in dimensions higher than 1, but with a strength
reduced from the results in 1D. 
In general, a given spin of an
hypercubic lattice would tend to form singlets 
with ${\rm 2\times D}$ partners in 
the absence of a vacancy and with ${\rm 2\times D-1}$ 
partners when a vacancy is
introduced next to it (as exemplified in Fig.2c for D=2).
The ideal enhancement factor F  would be
 ${\rm F \sim (2\times D)/(2\times D-1)}$, but it is certainly
reduced by the inclusion
of corrections beyond NN singlets. 
In Fig.4a, $C_S(i,j)$
for a (tilted) square cluster of 26 sites is shown with
and without a Zn-impurity. The relative enhancement at distance 1 along an
axis away from the vacancy is $\sim 11\%$, i.e. smaller than 
in 1D, as predicted. 
Our picture does not depend on the presence of long-range order,
as exemplified by calculations of
$C_S(i,j)$ for the frustrated $J_1-J_2$ 2D Heisenberg
model with a vacancy. Fig.4b actually
shows an AF enhancement near the vacancy $larger$ 
at $J_2/J_1=0.5$ (where the long-range AF order is suppressed)
than at $J_2/J_1=0.0$  ($\sim 31\%$
vs. $\sim 11\%$ for the correlation at distance 1).
A smooth connection between
the two results was observed numerically, and thus once again
the local AF enhancement seems caused by the RVB character of the
short distance fluctuations 
independently of the long distance
properties of the model.

\begin{figure}[htbp]
\centerline{\psfig{figure=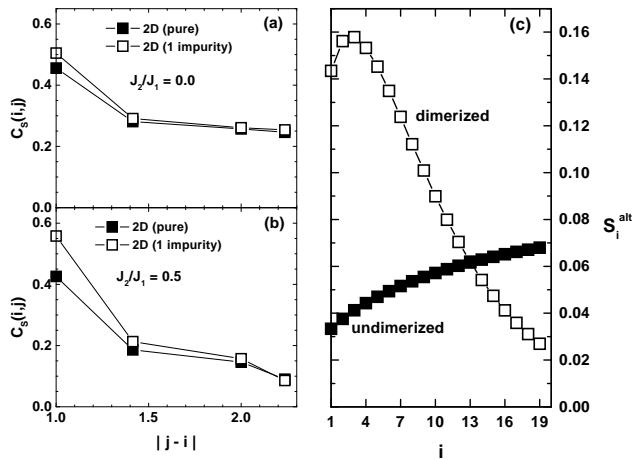,height=7cm}}
\vspace{0.5cm}
\caption{(a) $C_S(i,j)$ on a 26-sites 
tilted square cluster calculated with ED. Open (full)
squares denote results with (without) a vacancy. $i$ is
the site next to the vacancy if present. Correlations at distances 
$1, \protect\sqrt{2}, 2$ and $\protect\sqrt{5}$ away from the 
impurity are shown;
(b) Same as (a) but introducing frustration $J_2/J_1=0.5$;
(c) Alternating contribution to the {\it local} spin correlations
$S_i$ (see text and Ref. \protect\cite{eggert}) 
for a $J_1-J_2-\delta$ model with $J_2/J_1=0.2$
and $\delta = 0.03$ using DMRG on a chain of 160 sites with OBC
(open squares).
The full squares are results for $J_2=\delta = 0.0$ on the same chain.
In both cases 16 states are kept in the DMRG method, and the end of 
the chain is at site 0.}
\end{figure}

It is also possible to predict the
strength of the enhancement for models with spins $S > 1/2$.
Consider first a 1D chain. In the limit of $S \rightarrow \infty$, the
zero temperature ground state is antiferromagnetically ordered 
and the picture based on RVB singlets is no longer useful.
In this limit the presence of a nonmagnetic impurity is irrelevant for
the spin correlations which remain of maximum strength.
Assuming a monotonous crossover from $S=1/2$ to $S=\infty$
a Zn-induced enhancement of $C_S(i,j)$ for all $S$ would be expected, but
of  decreasing magnitude as $S$ grows. Indeed for a $S=1$ chain the
spin correlation at distance one 
near an end is $\sim 16\%$ larger than in  the bulk\cite{sorensen},
an enhancement 
smaller than for $S=1/2$\cite{foot1}. 
Thus, the vacancy-induced antiferromagnetic enhancement
seems to occur in any dimension D
and spin $S$, but the effect is the largest
for 1D $S=1/2$ systems where quantum fluctuations are crucial.

The results obtained in this paper allow us to make predictions for the NMR
spectra of a variety of compounds
similarly as done before for 1D $S=1/2$ chains\cite{eggert,takigawa}.
In Fig.4c the {\it alternating} part of the local spin correlations $S_i = \sum_j \langle 
S^z_i S^z_j \rangle$ (proportional to the local susceptibility $\chi_i$)
is shown. Here $\langle \rangle$ represents the expectation value in the lowest energy
state of the subspace of spin one, which would contribute to $\chi_i$ as
the temperature or an external magnetic field is increased from zero 
($S_i$ in the singlet ground state of a finite
lattice would trivially vanish). The figure shows that both dimerized and
undimerized chains have a nonzero local staggered magnetization near
vacancies. In the former the enhancement is concentrated in the vicinity
of the impurity, 
while in the latter it is spread over all the chain due to
spin-charge separation in excellent agreement with Ref.\cite{eggert}. 
NMR experiments should detect a broadening in
their spectra associated with this effect in several compounds such as
 CuGeO$_3$.
At a more speculative level, the AF enhancement discussed in this paper
may be responsible for the transition to
a 3D N\'eel order state in Zn-doped dimerized chains and
ladders\cite{recent1d,recentladder} .
Note, however, that this effect 
may depend on the strength of the (small)
coupling between the structures discussed here, and thus may 
vary appreciably from compound to compound.

We thank S. Eggert, I. Affleck, and N. Furukawa for useful
discussions.
E.D. is supported by grant NSF-DMR-9520776. 
G.B.M. acknowledges the financial support of the Conselho
Nacional de Desenvolvimento Cient\'{\i}fico e Tecnol\'{o}gico (CNPq-Brazil).


\begin{references}
\bibitem{comment2} Actually the 1-2 singlet is more probable
than 1-3 and 1-4  due to the characteristics of the 2-leg ladder
ground state\cite{ladder}. 
This detail is not qualitatively  relevant here.
\end{references}
\end{document}